\documentclass[aps,prb,twocolumn,sd,amsmath,amssymb,final,superscriptaddress]{revtex4-1}
\usepackage{dcolumn}% Align table columns on decimal point
\usepackage{graphicx}% Include figure files
\usepackage{epstopdf}
\usepackage{amsmath}
\usepackage[bookmarks=false,pdfstartview=FitH]{hyperref}
\usepackage{color}
\usepackage{booktabs}
\usepackage[all]{hypcap}
\newcommand\T{\rule{0pt}{2.6ex}}       % Top strut
\newcommand\B{\rule[-1.2ex]{0pt}{0pt}} % Bottom strut

\definecolor{orange}{cmyk}{0,0.4,0.8,0.2}
\definecolor{darkorange}{rgb}{.71,0.21,0.01}
\definecolor{darkgreen}{rgb}{.12,.54,.11}
\definecolor{darkblue}{rgb}{0.1,0.1,0.8}
\hypersetup{pdftex,
  breaklinks=true,  % so long urls are correctly broken across lines
  colorlinks=true,
  urlcolor=blue,
  linkcolor=blue,
  citecolor=blue,
  }

\def\Let@{\def\\{\notag\math@cr}}
\begin{document}

%\preprint{}

\title{Defect properties of Sn- and Ge-doped ZnTe: suitability for intermediate-band solar cells}


\author{Mauricio A. Flores}
\email[]{mauricio.flores@ug.uchile.cl}
%\thanks{}
\affiliation{Pontificia Universidad de Cat\'olica de Valpara\'iso, Casilla 4950, Valpara\'iso, Chile.}
\affiliation{Facultad de Ingenier\'ia y Tecnolog\'ia, Universidad San Sebasti\'an, Bellavista 7, Santiago 8420524, Chile.}


\begin{abstract}
We investigate the electronic structure and defect properties of Sn- and Ge- doped ZnTe by first-principles calculations within the DFT+\textit{GW }formalism. We find that $(\text{Sn}_\text{Zn})$ and $(\text{Ge}_\text{Zn})$ introduce isolated energy levels deep in the band gap of ZnTe, derived from Sn-$5s$ and Ge-$4s$ states, respectively. Moreover, the incorporation of Sn and Ge on the Zn site is favored in \emph{p}-type ZnTe, in both Zn-rich and Te-rich environments. The optical absorption spectra obtained by solving the Bethe-Salpeter equation reveals that sub-bandgap absorptance is greatly enhanced due to the formation of the intermediate band. Our results suggest that Sn- and Ge-doped ZnTe would be a suitable material for the development of intermediate-band solar cells, which have the potential to achieve efficiencies beyond the single-junction limit.
\end{abstract}

\date{\today}

%\pacs{}
%\keywords{}

\maketitle

\section{Introduction}

Zinc telluride (ZnTe) is a wide-gap II-IV semiconductor with a direct band gap of $\sim$2.26 eV at room-temperature. It is a potential candidate for solid-state applications such as green light emitting diodes (LEDs), \cite{Shan02,Tanaka09} terahertz (THz) imaging applications, \cite{Guo07} detectors, \cite{Wu96,Schall00} transparent conductors, \cite{Feng1996} and solar cells. \cite{Wu15,Wolden16} Moreover, ZnTe can be doped both \textit{p}- and \textit{n}-type, but the latter is difficult to achieve in practice due to the high concentration of native acceptors, such as (Te$_\text{i}$) and (V$_\text{Zn}$), that shift the position of the Fermi level toward the valence band, favoring \emph{$p$}-type conduction. \cite{Olusola16}

In semiconductor-based solar cells, one of the major limiting factor on the conversion efficiency is the incomplete utilization of the photon energy. Only photons whose energies are higher than to the energy difference between the bottom of the conduction band and the top of the valence band, i.e. the energy gap, can be absorbed to generate electron-hole pairs. Moreover, photo-generated carriers with energies in excess of the band gap are lost to heat as they rapidly thermalize; thus, the smaller the energy gap, more of the sun energy can be utilized. However, the largest recoverable voltage, i.e. the open-circuit voltage, is limited by the energy gap potential difference and decreases with decreasing the band gap of the semiconductor. In 1961, Shockley and Queisser \cite{Shockley61,Araujo94} found that the maximum efficiency for an ideal device with an energy gap of 1.1 eV (in which all recombination is assumed to be radiative) illuminated by black body radiation at 6000 K is 30.0\%. This result was extended to any absorption spectrum by Mathers,\cite{Mathers77} who found a limit of 31\% for the conversion efficiency of an ideal solar cell under AM1 spectrum.

Several strategies for increase the efficiencies beyond the Shockley-Queisser limit have been proposed in the last years. \cite{Green17} Improved photovoltaic conversion efficiencies can be achieved by using a sequence of materials of decreasing band gap such that each material absorbs in one part of the solar spectrum. \cite{King12} But, this approach is currently limited to concentrator \cite{Green15} and space systems due to its high manufacturing costs. Another suggestion,\cite{Wolf60} consists in the introduction of an isolated metallic band in the forbidden gap of a wide-gap semiconductor. This intermediate band (IB) allows additional optical transitions, thereby enabling sub-band gap energy photons to contribute to the photocurrent by pumping electrons from the valence band (VB) to the IB and from the IB to the conduction band (CB).\cite{Luque10,Luque12} Different approaches have been explored to implement the intermediate-band solar cell (IBSC) concept: (1) the use of highly mismatched alloys (HMAs), \cite{Lopez11,Ahsan12} a class of materials in which an isolated band is formed as a result of a band anti-crossing mechanism between the localized states of an isovalent dopant and the extended states of the host; \cite{Shan99,Wu02} (2) the development of quantum dots solar cells (QDSCs) in which a periodic array of quantum dots introduce an IB in the fundamental gap of a suitable host;\cite{Marti06,Shoji13,Yang13} (3) the use of heavily doped semiconductors \cite{Castan13,Hashemi14,Koskelo16,chen17,Huang17} in which a suitable dopant introduces its $d$ or $s$ orbitals deep in the band gap, giving rise to a delocalized impurity-band. \cite{Luque06,Luque10}

Due to its wide band gap, ZnTe has been proposed as a good candidate for the development of intermediate-band photovoltaic devices. For instance, ZnTe:O\cite{Lee10,Antolin14,Sharma17,Tanaka17} has been extensively investigated as a highly mismatched alloy with promising results reported in Refs. [\onlinecite{Wang09,Tanaka11,Tanaka13}]. However, \emph{n}-type doping may be required in order to partially fill the IB, so that two-photon photocurrent would be maximized. In addition, heavily doping\cite{Tablero06,Lee17} and co-alloying\cite{Kim14} were also proposed to create an intermediate band in bulk ZnTe. 

In this work, we investigate the role of Sn and Ge as impurities in ZnTe. We calculate their formation energies and charge transition levels within the DFT$\hspace{0.05cm}+\hspace{0.05cm}$\emph{GW} method, \cite{Hedstrom06,Rinke09,Flores16_1,Flores16_2} which combines quasiparticle energies obtained within the \emph{GW} approximation with total energy calculations based on the density functional theory (DFT). We find that under Zn-rich growth conditions the compensation mechanisms in Sn- and Ge- doped ZnTe are favorable for the formation of an isolated and half-filled intermediate-band, greatly enhancing the solar energy conversion efficiency by enabling the absorption of sub-bandgap photons in a two-step excitation process.

\section{Methods}

\subsection{Computational Details}

We performed total energy DFT calculations using the gradient-corrected exchange and correlation functional of Perdew, Burke, and Ernzerhof (PBE), \cite{Perdew96} as implemented in the Quantum-ESPRESSO package. \cite{Giannozzi2009} The electron-ion interactions were described by GBRV ultrasoft pseudopotentials. \cite{Garrity2014} We used a 36 Ry energy cutoff for the plane-wave basis set expansion and a 200 Ry cutoff to represent the charge density. Moreover, our calculations were performed using large 512-atom supercells, in which all the atoms were allowed to relax until the forces acting on each ion were smaller than 0.001 Ry/bohr. The Brillouin zone is sampled by the $\Gamma$ point only.

Many-body $G_0W_0$ calculations of defect-containing supercells were performed using the WEST code, \cite{Pham13,Govoni15} which implements the formalism proposed in Refs. [\onlinecite{Nguyen12}] and [\onlinecite{Pham13}] that avoids the explicit summation over empty electronic states by using a technique called projective eigendecomposition of the dielectric screening. In our calculations, we used 512 projective dielectric eigenpotential basis vectors to represent the dielectric matrix and norm-conserving Vanderbilt pseudopotentials (ONCV) \cite{Hamann13} including 20 and 16 valence electrons for Zn and Te atoms, respectively, with a plane-wave energy cutoff of 70 Ry. For the absolute position of the VBM, we used $\Delta E_\text{VBM} = -0.81$ eV, as obtained in Ref. [\onlinecite{Gruneis14}] employing the self-consistent \emph{GW}$\Gamma$ approximation, which includes a first-order vertex correction in the self-energy and the effect of the spin-orbit coupling as a \emph{posteriori} correction.

In addition, the optical properties were investigated within the \emph{GW}-BSE formalism, using the ABINIT code. \cite{Torrent08,Gonze2009,Gonze16}
We used a $3\times3\times2$ supercell of ZnTe containing a single substitutional Sn and Ge impurity (occupying the Zn site). The matrix elements of the Bethe-Salpeter (BSE) Hamiltonian were first calculated on a $3\times3\times4$ $\textbf{k}$-grid shifted along the (0.11, 0.12, 0.13) direction and subsequently interpolated onto a much finer $6\times6\times8$ grid, by using the technique proposed in Ref. [\onlinecite{Gillet16}]. We used the Tamm-Dancoff approximation \cite{Benedict98} in which only the resonant part of the BSE Hamiltonian is considered.

\subsection{Defect formation energies and chemical potentials}

The formation energy of a defect in charge state $q$ can be expressed as \cite{Jain11, Flores16_1,Flores17_2}
\begin{eqnarray}
E^f_q[\textbf{R}] = E_q[\textbf{R}] - E_\text{ref} + qE_F,
\label{ec:1}
\end{eqnarray}
\begin{eqnarray}
E_\text{ref} \equiv E^\text{ZnTe}_\text{bulk} + \sum_i n_i(\Delta\mu_i + \mu^\text{ref}_i),\label{ec:2}
\end{eqnarray}
where $E_q[\textbf{R}]$ is the total energy of the system in charge state $q$ and ionic configuration $\textbf{R}$, $E_\text{ref}$ is the energy of a reference system, i.e. the defect-free supercell, and $\text{E}_\text{F}$ is the Fermi energy. The integer $n_i$ corresponds to the number of atoms of species $i$ that are either added ($n_i > 0$) or removed ($n_i < 0$) from the reference system. $\Delta\mu_i$ is a relative chemical potential for the $i$th atomic species referenced to $\mu^\text{ref}_i$, which is
the chemical potential of its pure elemental phase, e.g., Zn (hexagonal with space group $P63$), Te (trigonal structure with space group $P3_121$), $\alpha$-Sn (cubic structure with $Fd$-3 space group), and $\alpha$-Ge (cubic structure with $Fd$-3 space group).

To ensure the stability of the ZnTe crystal, the chemical potentials must be thermodynamically limited by the following equation:
\begin{equation} \Delta\mu_{\text{Zn}} + \Delta\mu_{\text{Te}} = E^f[\text{ZnTe}], \label{ec:3} \end{equation}
where $E^f[\text{ZnTe}] = -0.92$ eV is the calculated formation enthalpy of bulk ZnTe. Moreover, since $\Delta\mu_{\text{i}} = 0$ means that the $i$th element is rich enough to form a pure solid phase, $\Delta\mu_{\text{Zn}} < 0$, $\Delta\mu_{\text{Te}} < 0$, $\Delta\mu_{\text{Ge}} < 0$, and $\Delta\mu_{\text{Sn}} < 0$ are also required. In addition, to avoid the formation of secondary phases of Ge or Sn with the host atoms, the chemical potentials are also bounded by the following relations:
\begin{equation} \Delta\mu_{\text{Ge}} + \Delta\mu_{\text{Te}} \leq E^f[\text{GeTe}] = -0.15 \text{ eV}, \label{ec:4} \end{equation}
\begin{equation} \Delta\mu_{\text{Sn}} + \Delta\mu_{\text{Te}} \leq E^f[\text{SnTe}] = -0.58 \text{ eV}, \label{ec:5} \end{equation}
where $E^f[\text{GeTe}]$ and $E^f[\text{SnTe}]$ are the calculated formation energies of GeTe and SnTe, respectively. In the case of ZnTe:Ge, considering Eq. (\ref{ec:4}) and the need of avoid the formation of pure phases of germanium, the chemical potential of Ge is restricted by

\begin{equation} \Delta\mu_{\text{Ge}} \leq \min(0,E^f[\text{GeTe}]- \Delta\mu_{\text{Te}}).\label{ec:6}\end{equation}

Hence, under Te-rich conditions we have $\Delta\mu_{\text{Te}} = 0$ and $\Delta\mu_{\text{Zn}} = -0.92$, then
\begin{equation} \Delta\mu^\text{Te-rich}_{\text{Ge}} \leq -0.15 \text{ eV}.\label{ec:7} \end{equation}

For Zn-rich conditions, $\Delta\mu_{\text{Te}} = -0.92$ and $\Delta\mu_{\text{Zn}} = 0$, then
\begin{equation} \Delta\mu^\text{Zn-rich}_{\text{Ge}} \leq 0 \text{ eV}.\label{ec:8} \end{equation}

Similarly, for the case of ZnTe:Sn, the chemical potential of tin is bounded by

\begin{equation} \Delta\mu_{\text{Sn}} \leq \min(0,E^f[\text{SnTe}]- \Delta\mu_{\text{Te}}).\label{ec:9}\end{equation}

Under Te-rich conditions, $\Delta\mu_{\text{Te}} = 0$ and $\Delta\mu_{\text{Zn}} = -0.92$, then
\begin{equation} \Delta\mu^\text{Te-rich}_{\text{Ge}} \leq -0.58 \text{ eV}.\label{ec:10} \end{equation}

And, for Zn-rich conditions $\Delta\mu_{\text{Te}} = -0.92$ and $\Delta\mu_{\text{Zn}} = 0$, then
\begin{equation} \Delta\mu^\text{Zn-rich}_{\text{Ge}} \leq 0 \text{ eV}.\label{ec:11} \end{equation}

\subsection{DFT$\hspace{0.05cm}+\hspace{0.05cm}$GW formalism}
According to Eq. (\ref{ec:1}), the formation energy of a defect in charge state $q\hspace{-0.05cm}-\hspace{-0.07cm}1$ is given by
\begin{eqnarray}
E^f_{q-1}[\textbf{R}_{q-1}] = E_{q-1}[\textbf{R}_{q-1}] - E_\text{ref} + (q-1)E_F.\label{ec:12}
\end{eqnarray}
Adding and subtracting first $E_{q-1}[\textbf{R}_{q}]$ and then $E_{q}[\textbf{R}_{q}]$, we have
\begin{equation}
\begin{split}
E^f_{q-1}[\textbf{R}_{q-1}] &= \left \{ E_{q-1}[\textbf{R}_{q}] - E_{q}[\textbf{R}_{q}] \right \} \\
&+ \left \{ E_{q-1}[\textbf{R}_{q-1}] - E_{q-1}[\textbf{R}_{q}]\right \} \\
&+ E^f_{q}[\textbf{R}_{q}] - E_F \\
&\equiv E_\text{QP} + E_{\text{relax}} + E^f_{q}[\textbf{R}_{q}] - E_F.\label{ec:13}
\end{split}
\end{equation}

The first term in the last equation, $E_\text{QP} = \left \{ E_{q-1}[\textbf{R}_{q}] - E_{q}[\textbf{R}_{q}] \right \}$, is a charged excitation, i.e. an electron addition or electron removal energy. This quantity is usually not well described within DFT, but it may be evaluated using the \emph{GW} approximation, which can provide an accurate description of excited states.\cite{Hedin65} The second term, $E_{\text{relax}}$ = $ E_{q-1}[\textbf{R}_{q-1}] - E_{q-1}[\textbf{R}_{q}]$, corresponds to a structural relaxation energy that could be evaluated at DFT-level, since we avoid the computation of energy differences between configurations with distinct number of electrons. \\
Similarly, we obtain
\begin{equation}
\begin{split}
E^f_{q+1}[\textbf{R}_{q+1}] &= \left \{E_{q+1}[\textbf{R}_{q}] - E_{q}[\textbf{R}_{q}]\right \} \\
&+ \left \{ E_{q+1}[\textbf{R}_{q+1}] - E_{q+1}[\textbf{R}_{q}]\right \} \\
&+ E^f_{q}[\textbf{R}_{q}] + E_F \\
&\equiv E_\text{QP} + E_{\text{relax}} + E^f_{q}[\textbf{R}_{q}] + E_F.\label{ec:14}
\end{split}
\end{equation}

Using Kohn-Sham (KS) eigenvalues $\epsilon_{n,k}^{\text{KS}} $ and wave functions $\psi_{n,k}^{\text{KS}} $ as mean-field starting point, the quasiparticle energy is calculated by adding the first-order perturbative correction
\begin{equation}
E^\text{QP}_{n,k} = \epsilon_{n,k}^{\text{KS}}  + \left< \psi_{n,k}^{\text{KS}} |\Sigma (E^{\text{QP}}_{n,k}) -  V_\text{xc}| \psi_{n,k}^{\text{KS}}\right>,\label{ec:15}
\end{equation}
which comes from replacing the KS exchange-correlation potential $V_{xc}$ with the self-energy operator $\Sigma$, which contains the effects of the exchange and correlation among the electrons.

Additionally, it is important to note that quasiparticle corrections obtained within the $G_0W_0$ approximation mainly reflect the difference between $V_{xc}$ and the non-local electron self-energy operator $\Sigma$, and thus are less dependent to the supercell size. \cite{Choi09} Therefore, considering the high computational demands, in the present work we calculated the quasiparticle $G_0W_0$ corrections by using 64-atom defect-containing supercells at the $\Gamma$ point only. These corrections were then used to correct the KS eigenvalues of 512-atom supercells by means of a scissors operator, and obtain the quasiparticle energies of interest with respect to the average electrostatic potential.

\subsection{Expected level of accuracy}

In supercell calculations, when periodic boundary conditions are imposed, finite-size errors arise due to spurious interactions between the periodic images of defects. The structural distortions around the defects may result in long-range elastic forces that may affect the calculation of relaxation energies within the DFT+\textit{GW} scheme. Moreover, in the case of charged systems, the electrostatic error in the individual DFT eigenvalues needs to be accounted for.\cite{Jain11}

\begin{table}[h]
\caption{\label{tab:t1}Defect formation energies used as reference for the DFT$\hspace{0.05cm}+\hspace{0.05cm}$\emph{GW} scheme. The Fermi energy is set to the valence band maximum (all values are given in eV).}
\begin{ruledtabular}
\vspace{0.1cm}
\begin{tabular}{ccc}
 \hspace{0.5cm}\text{Reference system} & \hspace{0.4cm}$E^f$ (Te-rich) & $E^f$ (Zn-rich) \B \\
\hline
\T  \T \hspace{0.5cm}$(\text {Sn}_\text{Cd})^{+2}$ & \hspace{0.15cm}$-0.72$  & \hspace{-0.2cm}$-0.39$\\
\T \hspace{0.5cm}$(\text{Sn}_{\text{Zn}}\hspace{-0.05cm}-\hspace{-0.05cm}\text{V}_\text{Zn})^{0}$ & \hspace{0.4cm}$1.55$  & $2.77$\\
\T \hspace{0.5cm}$(\text {Sn}_\text{i})^{+2}$ & \hspace{0.4cm}0.64  & 0.05\\
\T \hspace{0.5cm}$(\text {Sn}_\text{Te})^0$ & \hspace{0.4cm}3.86 & 2.35 \\\vspace{-0.25cm}\\
\hline
\T \hspace{0.5cm}$(\text {Ge}_\text{Cd})^{+2}$ & \hspace{0.15cm}$-0.79$  & \hspace{-0.2cm}$-0.02$\\
\T \hspace{0.5cm}$(\text{Ge}_{\text{Zn}}\hspace{-0.05cm}-\hspace{-0.05cm}\text{V}_\text{Zn})^{0}$ & \hspace{0.4cm}$1.45$  & $3.14$\\
\T \hspace{0.5cm}$(\text {Ge}_\text{i})^{+2}$ & \hspace{0.4cm}0.19  & 0.04\\
\T \hspace{0.5cm}$(\text {Ge}_\text{Te})^0$ & \hspace{0.4cm}3.42 & 2.35 \\
\end{tabular}
\end{ruledtabular}
\end{table}

To avoid the finite-size effects as much as possible, we calculated both relaxation and excitation energies by using 512-atom supercells. The elastic effects are expected to be negligible small. Moreover, due to the high dielectric constant of ZnTe, the position of the defect levels relative to the valence band maximum are within the numerical accuracy of the $G_0W_0$ calculations (about $0.10-0.15$ eV). The calculated defect formation energies obtained within DFT that were used as reference for the DFT+\textit{GW} scheme are shown in Table \ref{tab:t1}.

\section{Results and discussion}

\subsection{Defect formation energies}

We first calculate the formation energies of Sn impurities in ZnTe. We consider tin atoms occupying substitutional sites, i.e. $(\text{Sn}_\text{Zn})$ and $(\text{Sn}_\text{Te})$; the interstitial site $(\text{Sn}_\text{i})$; and the possible formation of a defect complex of Sn with a Zn vacancy, i.e. the $(\text{Sn}_\text{Zn}-\text{V}_\text{Zn})$ complex.

\begin{figure}[!h]%
\vspace{0.2cm}
 \centering
 \includegraphics[width=8.5cm]{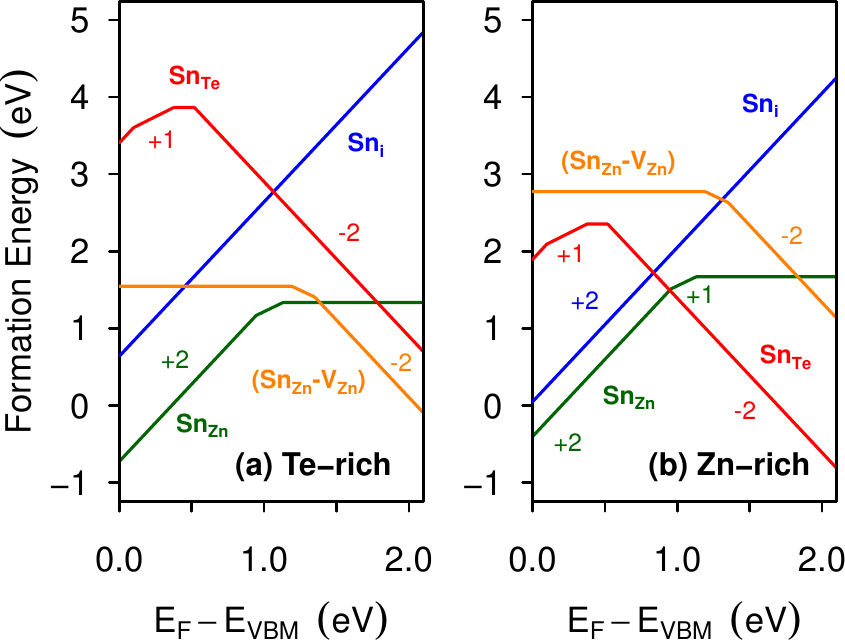}
 \caption{Defect formation energies of Sn impurities in ZnTe as a function of the Fermi-level position within the band gap, under (a) Te-rich growth conditions and (b) Zn-rich growth conditions.}
 \label{fig:1}
\end{figure}

\begin{figure}[h]%
 \centering
 \includegraphics[width=3.75cm]{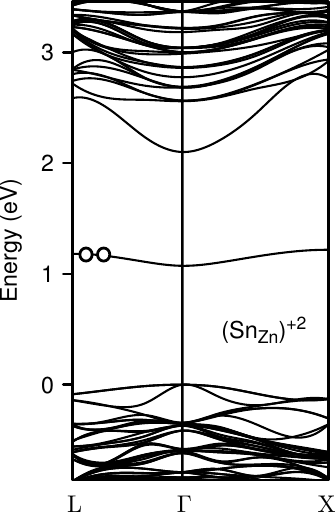}
  \hspace{-0.1cm}
  \includegraphics[width=4.5cm]{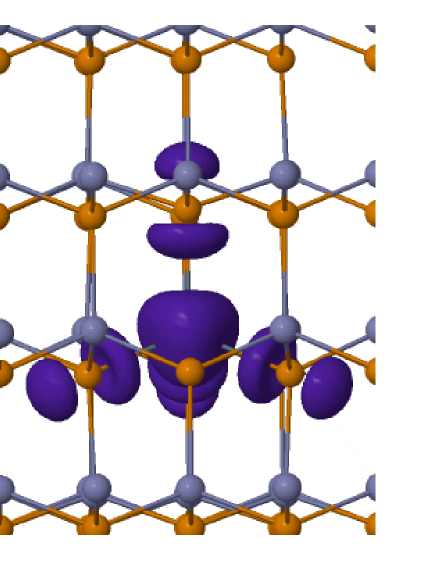}
 \caption{Electronic band structure of $(\text{Sn}_\text{Zn})^{+2}$ and the charge density isosurface ($\rho = 0.0005e/\text{Bohr}^3$) corresponding to the isolated energy level in the band gap. The calculations were performed in a 250-atom supercells and $G_0W_0$ corrections were considered through the application of a scissors operator at the $\Gamma$ point. The empty circles indicate the occupation of the energy level in the band gap.}
\label{fig:2}
\end{figure}

Figure \ref{fig:1} shows the calculated formation energies under Te-rich and Zn-rich growth conditions. In all cases, with the sole exception of $(\text{Sn}_\text{i})$, we observe that the incorporation of Sn creates deep charge transition levels in the band gap. Moreover, when the position of the Fermi energy is near to mid-gap or close to the VBM, we find that the most favorable is the substitutional site $(\text{Sn}_\text{Zn})$, in which Sn acts as a donor. According to our calculations, $(\text{Sn}_\text{Zn})$ introduces two charge transition levels $\epsilon (+2/+)$ and $\epsilon (+/0)$ at VBM+0.94 eV and VBM+1.12 eV, respectively. This result differs from the analogous $(\text{Sn}_\text{Cd})$ in CdTe. It was found that the latter exhibits a negative-U behaviour, i.e. $\epsilon (+2/0)$ is lower in energy than $\epsilon (+2/+)$.\cite{Babentsov07,Flores17_1} The absence of negative-U effect in $(\text{Sn}_\text{Zn})$ may be due to the larger band gap of ZnTe that helps to stabilize the $(\text{Sn}_\text{Zn})^{+1}$ configuration, which has an unpaired electron occupying an energy level deep in the gap (mostly derived from Sn $5s$ states). The calculated band structure of $(\text{Sn}_\text{Zn})^{+2}$  and the charge density isosurface corresponding to the isolated energy level in the band gap are shown in  Figure \ref{fig:2}.

When the Fermi level is close to the conduction band, we find that the most stable site depends on the chemical potential. Under Te-rich conditions, the formation of $(\text{Sn}_\text{Zn}-\text{V}_\text{Zn})$ complexes is preferred. This complex is a deep acceptor that introduces an isolated energy level in the gap mainly derived from Sn $5s$ states. It introduces two charge transition levels $\epsilon(0/-)$ and $\epsilon(-/{-2})$ at VBM$+1.20$ eV and VBM$+1.36$ eV, respectively. Under Zn-rich conditions, the substitutional $(\text{Sn}_\text{Te})$ is the dominant defect. It acts as deep acceptor introducing four charge transition levels at VBM$+0.1$ eV for $\epsilon({+2}/+)$, VBM$+0.36$ eV for $\epsilon(+/0)$, VBM$+0.56$ eV for $\epsilon(0/-)$, and VBM$+0.48$ eV for $\epsilon(-/{-2})$. In T$_d$ symmetry, the $(\text{Sn}_\text{Te})$ introduces a three-fold degenerate $t_2$ level deep in the band gap, thus is subject to Jahn-Teller distortions. \cite{Opik57,Goodenough98} We found that the T$_{d}$ to C$_{3v}$ distortion splits the $t_2$ level into a two-fold degenerate $e$ state and a non-degenerate $a_1$ level that lies deep in the band gap. The latter level is likely to act as a non-radiative Shockley-Read-Hall recombination center \cite{Shockley52} with a deleterious impact in carrier transport. However, for values of the Fermi level above $\epsilon(-/{-2})$ the $t_2$ level is fully occupied and the Jahn-Teller distortion will not occur. The interstitial site $(\text{Sn}_\text{i})$ is a shallow donor, which does not introduce any charge transition level in the band gap. The Sn impurity occupies an octahedral coordinated position in which its four nearest neighbours are Zn atoms. Moreover, its formation energy is lower under the Zn-rich growth conditions.

\begin{figure}[h]%
 \centering
  \vspace{0.1cm}
 \includegraphics[width=8.5cm]{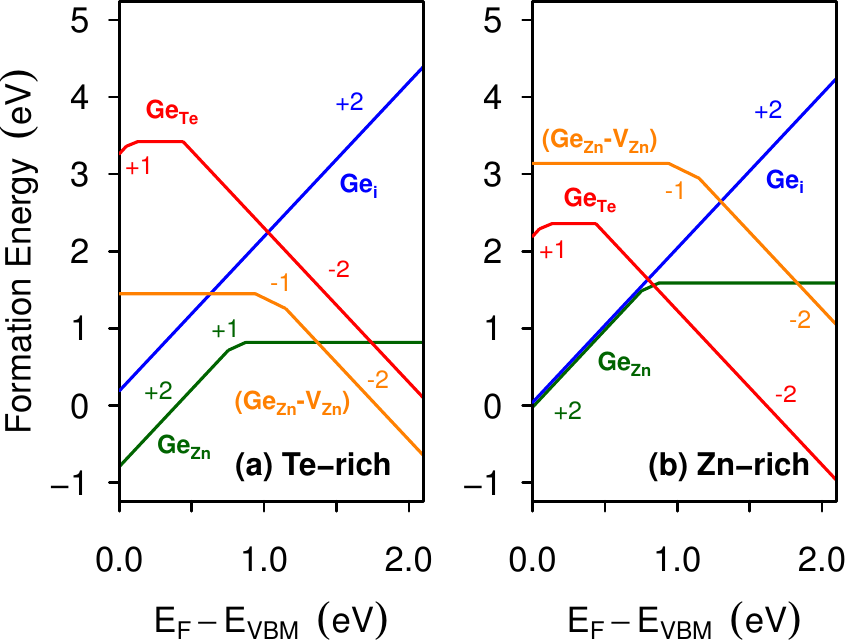}
 \caption{(Color online) Electronic band structures of (a) $(\text{Ge}_\text{Zn})^{+2}$, and (b) $(\text{Sn}_\text{Zn})^{+2}$, calculated by using 250-atom supercells. The empty circles indicate the occupation of the energy level in the band gap.}
  \label{fig:3}
\end{figure}

Next, we focus on the Ge doping. The calculated formation energies are shown in Figure \ref{fig:3}; we can see that, overall, our results are qualitative similar to those previously found for Sn. A major discrepancy occurs, however, in the case of $(\text{Ge}_\text{i})$ which is also a shallow donor, but the Ge impurity occupies an octahedral position surrounding by Te atoms instead of Zn atoms, as is the case of $(\text{Sn}_\text{i})$.

In Zn-rich and \emph{p}-type ZnTe, the formation energies of $(\text{Ge}_\text{i})^{+2}$ and $(\text{Ge}_\text{Zn})^{+2}$ are nearly degenerate, suggesting that the incorporation of Ge via interstitial diffusion is more favorable than Sn in this regime. Moreover, when the Fermi level is near the VBM, the substitutional $(\text{Ge}_\text{Zn})$ is the most stable site for both Te-rich and Zn-rich conditions. The electronic structure of this defect is similar to that of $(\text{Sn}_\text{Te})$, exhibiting an isolated level derived from Ge $4s$ states deep in the gap. It introduces two donor levels $\epsilon (+2/+)$ and $\epsilon (+/0)$ at 0.75 eV and 0.86 eV above the VBM, respectively. In the charge state $+2$, the Ge atom occupies the center of a Zn vacancy forming four equivalent Ge-Te bonds. The nearest-neighbor Te atoms relax slightly outwards by $\sim0.04$\%, resulting in Ge-Te bond lengths of 2.67 \AA; on the other hand, the $(\text{Sn}_\text{Zn})^{+2}$ configuration causes a more extend outward relaxation of $\sim 5.64$\%, resulting in a Sn-Te bond length of 2.81 \AA.

\begin{figure}[h]%
 \centering
 \hspace{-0.2cm}
 \includegraphics[width=3.75cm]{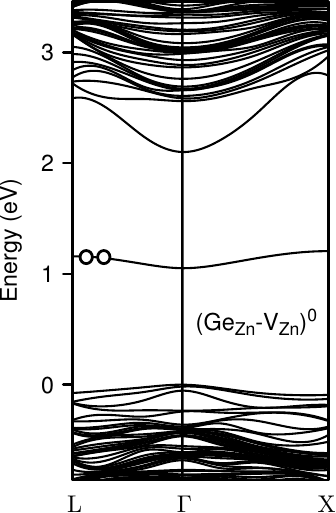}
   \hspace{-0.1cm}
  \includegraphics[width=4.7cm]{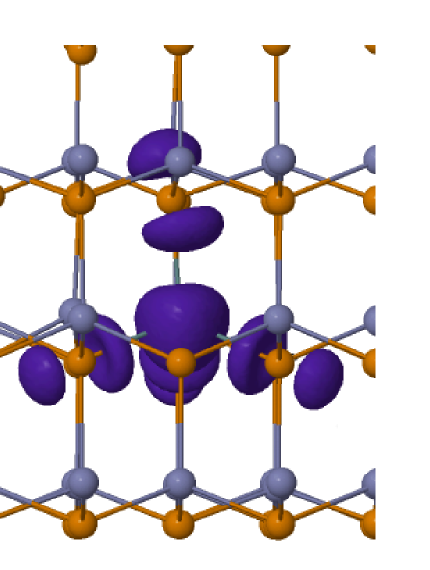}
 \caption{Electronic band structure of $(\text{Ge}_\text{Zn}-\text{V}_\text{Zn})^{0}$ and the charge density isosurface ($\rho = 0.0005e/\text{Bohr}^3$) corresponding to the isolated energy level in the band gap. The calculations were performed in a 250-atom supercells and $G_0W_0$ corrections were considered through the application of a scissors operator at the $\Gamma$ point. The empty circles indicate the occupation of the energy level in the band gap.}
  \label{fig:4}
\end{figure}

In addition, Ge impurities can form $(\text{Ge}_\text{Zn}-\text{V}_\text{Zn})$ complexes. As shown in Figure \ref{fig:4}, they introduce an isolated energy level in the band gap, which derives mainly from Ge $4s$ states. Moreover, these complexes are easier to form under Te-rich growth conditions since a high concentration of Zn vacancies is expected. They act as deep acceptors introducing two charge transition levels at  $\epsilon (0/-)$ and $\epsilon (-/{-2})$ at \text{VBM} + 0.95 eV and \text{VBM} + 1.16 eV, respectively. Finally, we find that the substitutional $(\text{Ge}_\text{Te})$ configuration in the charge state $q={-2}$ has T$_{d}$ symmetry. For the charge states $q = {-1}$, $0$, ${+1}$, and ${+2}$, it undergoes a T$_{d}$ to C$_{3v}$ distortion, similarly to $(\text{Sn}_\text{Te})$. It introduces a double-donor level $\epsilon ({+2}/+)$ at 0.05 eV, a single-donor level $\epsilon (+/0)$ at 0.11 eV, a single-acceptor level $\epsilon (0/-)$ at 0.55 eV, and a double-acceptor level $\epsilon (-/{-2})$ at 0.33 eV above the VBM.

\subsection{Suitability for intermediate-band solar cells}

In the previous section, we identified several defect configurations that introduce isolated energy levels in the fundamental gap of ZnTe, namely $(\text{X}_\text{Zn})$ and $(\text{X}_\text{Zn}-\text{V}_\text{Zn})$ (X = Sn, Ge). In the dilute limit, they can be thought as isolated defects surrounded by millions of atoms of the host material; thus, quantum-mechanical interactions between them are expected to be negligible small. Furthermore, if they are introduced at high concentrations, i.e. exceeding the Mott limit, \cite{Mott68} the atomic wave functions localized at different sites would overlap forming a delocalized impurity band, which can suppress the non-radiative recombination associated with impurity states.\cite{Olea10,Luque12,Flores17_1}
A delocalized impurity band avoids a strong electron-phonon coupling with the lattice (which facilitates non-radiative processes), as absorption and emission transitions do not involve delocalized-to-localized and localized-to-delocalized charge redistributions that could drive the system strongly out of equilibrium.

For an efficient operation of an IBSC, the IB must be half-filled and well isolated in the band gap. In this scenario, incoming photons are not only able to pump electrons from the VB to the CB, they also can pump electrons from the VB to the IB and from the IB to the CB, thereby allowing the absorption of low energy photons via a two-step excitation process. \cite{Wahnon02,Luque12} It is interesting to note that in the case of Sn- and Ge-doped ZnTe the half-filling of the IB can be naturally achieved under Zn grow conditions due to the amphoteric behaviour of the dopants. At high concentrations, the compensation between donors, $(\text{Sn}_\text{Zn})^{+2}$ and $(\text{Ge}_\text{Zn})^{+2}$, and acceptors, $(\text{Sn}_\text{Te})^{-2}$ and $(\text{Ge}_\text{Te})^{-2}$, will lead to the Fermi level pinning at the impurity band.

\subsection{Optical properties}

 \begin{figure}[h]
  \vspace{0.2cm}
  \centering
  \includegraphics[width=8cm]{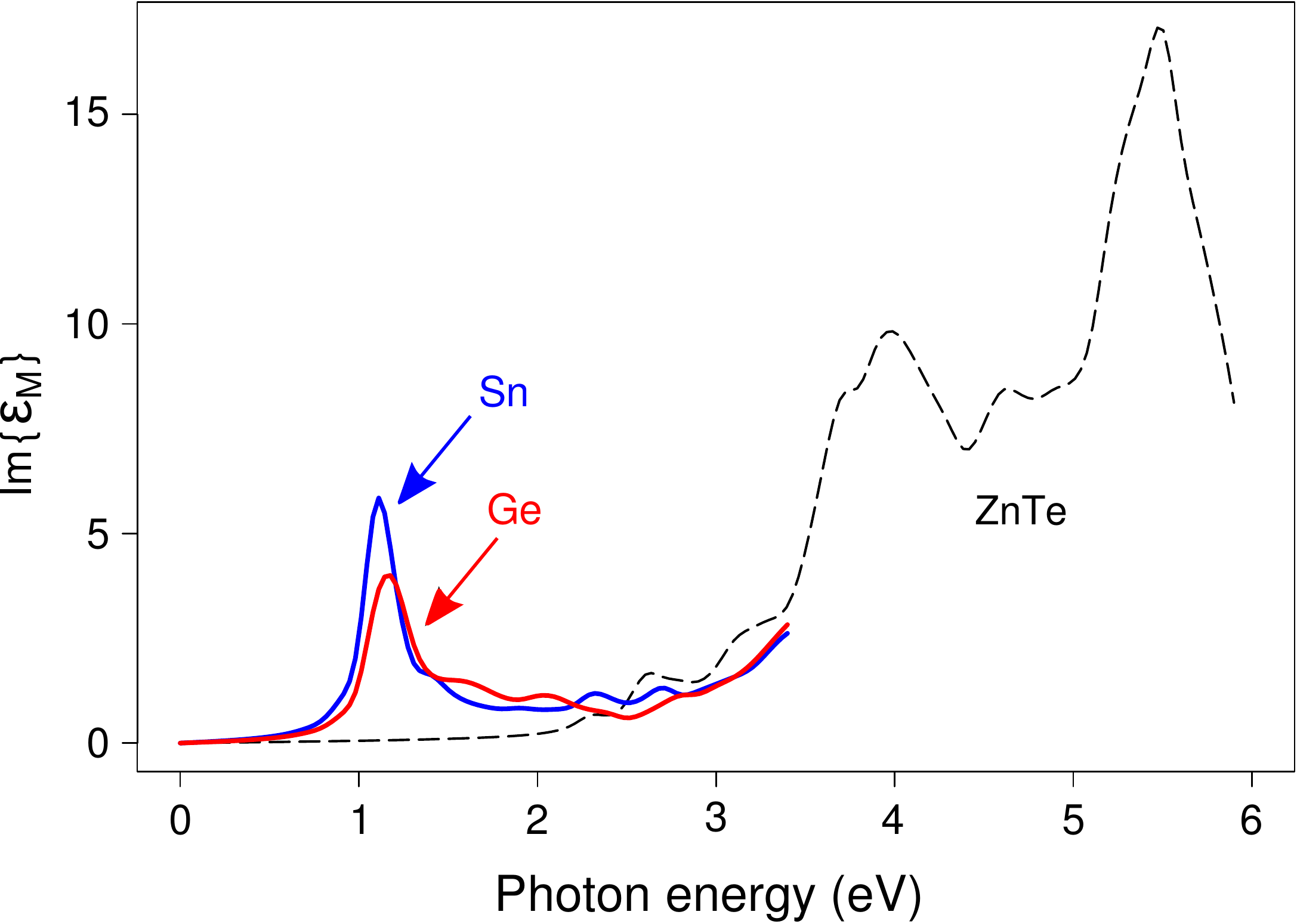}
  \caption{Imaginary part of the dielectric function for pristine ZnTe (dashed line) and ZnTe containing single substitutional Sn and Ge impurities occupying a Zn site (solid lines).}
   \label{fig:5}
 \end{figure}

Next, we investigate the optical spectra of pristine ZnTe and Sn- and Ge-doped ZnTe, given by the imaginary part of the macroscopic dielectric function $\text{Im}\{\epsilon_\text{M}(\omega)\}$. We performed first-principles calculations based on the \emph{GW}+BSE approach, which proceeds in the following three steps: (1) a ground-state DFT calculation; (2) a \emph{GW} calculation to correct the Kohn-Sham eigenvalues; and (3) the solution of the Bethe-Salpeter equation (BSE) \cite{Rohlfing98, Onida02} using the corrected eigenvalues to obtain the optical absorption spectra.

Figure  \ref{fig:5} shows the calculated $\text{Im}\{\epsilon_\text{M}(\omega)\}$ for pristine and doped ZnTe. We can see that both Sn and Ge doping allow the absorption of photons in the sub-band gap region. The absorption peaks near 1.1 eV are due to transitions between the intermediate-band and the conduction-band, whereas the absorption in the energy region up to the direct band gap of ZnTe is mainly due to transitions between the valence-band and the intermediate-band.
Our results indicate that the intermediate-band in Sn- and Ge-doped ZnTe acts as a stepping stone allowing the optical excitation of sub-bandgap photons.

\section{Summary}

In summary, we report a comprehensive theoretical study on the defect properties of Sn- and Ge-doped ZnTe. We used the DFT+\textit{GW} approximation and found that both $(\text{Sn}_\text{Zn})$ and $(\text{Ge}_\text{Zn})$ introduce well isolated intermediate bands in the fundamental gap of ZnTe. Moreover, the calculated absorption spectra indicates that the optical excitation of sub-bandgap photons is greatly enhanced by the presence of the IB. Our results suggest that ZnTe:Sn and ZnTe:Ge are suitable candidates for the development of high-efficiency
 IBSC devices.

\begin{acknowledgments}
Powered@NLHPC: This research was partially supported by the supercomputing infrastructure of the NLHPC (ECM-02).
\end{acknowledgments}

\bibliographystyle{apsrev4-1}
\bibliography{bib}

\end{document}